\documentclass[%
reprint,
superscriptaddress,
 amsmath,amssymb,
aps,
pra,
showkeys,
showpacs,
]{revtex4-2}

\usepackage{graphicx}

\begin{document}

\title{Scattering Angle Dependence of Fano Resonance Profiles in Cold Atomic Collisions Analyzed with the Complex Valued $ \boldsymbol{w} $ Parameter}

\author{Tanmay Singh}
\altaffiliation[Current address: ]{School of Earth and Space Exploration, Arizona State University, AZ 85281, Tempe, USA}
\affiliation{Department of Physics, Indian Institute of Technology Delhi, Hauz Khas, New Delhi 110016, India}

\author{Raj Aryan Singh}%
\altaffiliation[Current address: ]{Centre for Astrophysics and Supercomputing (CAS), Swinburne University of Technology, Hawthorn, Australia}
\affiliation{Department of Physics, Indian Institute of Technology Delhi, Hauz Khas, New Delhi 110016, India}

\author{Fumihiro Koike}
\email{koikef@sophia.ac.jp}
\affiliation{Department of Materials and Life Sciences, Sophia University, Tokyo, 102-8554, Japan}

\author{Masatomi Iizawa}
\altaffiliation[Current address: ]{Institut f\"ur Theoretische Physik, Technische Universit\"at Braunschweig, Mendelssohnstr. 3, D-38106 Braunschweig, Germany}
\affiliation{Department of Physics, Rikkyo University, Tokyo 171-8501, Japan}

\author{Yoshiro Azuma}
\email{y-azuma@sophia.ac.jp}
\altaffiliation[Current address: ]{Department of Physics, Birla Institute of Technology and Science—Pilani, K. K. Birla Goa Campus, NH-17B, Zuarinagar, Sancoale, Goa- 403726, India}
\affiliation{Department of Physics, Indian Institute of Technology Delhi, Hauz Khas, New Delhi 110016, India}
\affiliation{Department of Materials and Life Sciences, Sophia University, Tokyo, 102-8554, Japan}


\begin{abstract}
The scattering angle dependence of Fano resonance profiles in cold atomic collisions
has been theoretically studied.
A complex-valued parameter $ w $ with an analytical formula describing the asymmetry of the resonance profile is proposed 
as a new development from previous work on electron resonance scattering from atoms
[F. Koike, J. Phys. $\mathbf{B10}$, 2883 (1977)]. 
It serves as the general formulation which leads to the angle-dependence of Fano's $ q $ parameter.
Calculations for the case of cold elastic collisions of hydrogen atoms with krypton atoms 
have been accomplished. The strong angle dependence of the resonance profile asymmetry in the  
differential scattering cross section due to the interference
with non-resonant partial waves is demonstrated. 
The scattering angle dependence of the resonance profile and thus the newly proposed asymmetry parameter  
is highly sensitive to the inter-atomic interaction potentials. 
This is likely to prove useful for the study of interaction potentials themselves.
\end{abstract}


\maketitle


\section{\label{sec:Introduction}Introduction
}
The Fano resonances represented by the Fano's $q$ parameter have been extensively studied since its inception\cite{Fano_1961}. 
The concept was originally studied in atomic and molecular physics but subsequently found use in various phenomena including classical mechanics, electrical engineering, modern optics etc.\cite{Iizawa_2021}. Three tiers can be recognized for applying the concept in quantum mechanical resonance scattering studies depending on the energies and physical scale.
They are 1. nuclear collisions, 2. electron scattering and photoionization from atoms and molecules, and 3. the recently evolving area of cold atomic and molecular collisions. Wigner\cite{Wigner_PhysRev.70.15,Wigner_PhysRev.70.606} discussed
the nuclear reactions and proposed Wigner's one-level resonance formula. 
Gibson and Dolder\cite{Gibson_1969} observed the angle-dependent 
asymmetry of the resonance profiles in 
the differential elastic electron scattering from helium atoms.
Koike\cite{Koike_1977}
pointed out that this asymmetry
can be described by an angle-dependent complex valued $w$ parameter, and also that $w$ produces the angular dependence of the Fano's $q$ parameter.  

In general, a well-isolated quantum mechanical resonance can be observed only when the de Broglie wavelength of the collision system is comparable to the range of interaction between the colliding particles.
For atomic and molecular collisions, it is the ``cold collision" 
that satisfies such a condition.
The incident particle may be temporarily trapped when  
the system undergoes resonance scattering.
Such ``orbiting resonances",
that are characterized by the formation of a complex orbiting around the colliding partners,
were observed by  Toennies et al\cite{Toennies_1979}
in collisions of hydrogen atoms and hydrogen molecules with rare gas atoms 
at the center of mass collision energies ranging from 0.5 to 100 meV. 

Subsequently, such resonance phenomena in ``cold" atomic and molecular
collisions have been studied by a number of authors\cite{Chilcott_2022,Li_2021,Samuelis_2000,Lysebo_2009,Weckesser_2021,Blech_2020,Vogels_2018,Naidon_2019,Abeelen_1999,Paliwal_2021,Jongh_2020}. 
Recently, the scattering angle dependence of the resonance 
profiles for cold $\mathrm{He}^* + \mathrm{D}_2 $ collisions was experimentally measured at the energy of 4.8 Kelvin by
Paliwal et al\cite{Paliwal_2021}.
They discussed the asymmetry of the resonance profiles
that appeared in the energy spectra of the differential
scattering cross-sections in terms of 
Fano resonance\cite{Fano_1961}.
However, as discussed later in section~\ref{sec:Conclusion}, their data analysis failed to avoid inadequacies in view of the theoretical treatment of the resonance profiles. Consequently, it is crucial to establish the proper methodology for the resonance spectral analysis in cold atomic and molecular collisions.
The angle dependence of the asymmetry in the resonance profiles
gives a very sensitive tool to determine the inter-atomic
interaction potential which leads to van der Waals forces. 
In the present paper, we consider
the angle-dependent Fano-type asymmetry of the resonance profiles 
in cold atomic and molecular collisions
following the derivation given in ref~\cite{Koike_1977}.

Quantum mechanical resonances take place 
when a scattering wavefunction is pulled into the 
interaction region to form a quasi-bound state.
The scattering phase-shift changes by 
$ \pi $ radian when the collision energy passes through the resonance point.
The asymmetry of the profiles for an 
isolated resonance in elastic scattering is determined solely by 
the background phase shifts 
of the collision system\cite{Koike_1977}. 
In the present paper, we shall apply the formalism developed for electron scattering\cite{Koike_1977} 
to the resonance in cold
atomic and molecular collisions. An angle-dependent
resonance profile parameter $w$ that is analytic on the Gauss plane
is introduced. A generalized mathematical expression for the 
angle-dependent real-valued Fano profile parameter $q$ is given in terms of the complex-valued $w$ parameter.
The orbiting resonances in $\mathrm{H} + \mathrm{Kr}$ collisions
are discussed in detail using the inter-atomic interaction 
potentials that are proposed by Toennies et al.\cite{Toennies_1979}.
It is shown that the $w$ parameter describing the angle-dependent resonance profile
becomes a very stringent tool to analyze the inter-atomic interaction
potentials.

In the following section, we develop the theory of our complex $w$ parameter describing angle-dependent 
resonance profiles in elastic differential scattering. 
In section \ref{sec:H+Kr},
we theoretically calculate the actual case of $\mathrm{H} + \mathrm{Kr}$ collisions. 
In section \ref{sec:Discussion},
we discuss the characteristics of the calculated resonance profiles 
and develop the methodology of angle-dependent
resonance profile analysis.
Finally, in section \ref{sec:Conclusion}, we give concluding remarks. 

\section{\label{sec:Theory}Theory}
We consider the elastic scattering of particles in which
the effect of spin or the internal angular momentum 
of the colliding pair can be neglected.
According to Blatt and Biedenharn\cite{Blatt_1952}, 
the angle-differential scattering cross section $ {\mathrm{d}\sigma}/{\mathrm{d}\Omega} $ can be given by
\begin{equation}
\label{DifferentialCrossSection}
\frac{\mathrm{d}\sigma}{\mathrm{d}\Omega} = \frac{1}{k^2}\sum_{L=0}^\infty B_LP_L(\cos\theta)
\end{equation}
with
\begin{eqnarray}
\label{BL}
B_L = \sum_{\ell=0}^{\infty}~\sum_{\ell^\prime = |\ell-L|}^{\ell+L}
(2\ell+1)(2\ell^\prime+1)
[C(\ell,\ell^{\prime}L;00)]^2\nonumber \\
\times
\sin\eta_\ell\sin\eta_{\ell^\prime}\cos(\eta_\ell - \eta_{\ell^\prime}),
\end{eqnarray}
where $k$ is the wavenumber, $ \ell $ and 
$ \ell ^ \prime $ are angular momentum quantum numbers, $ \eta _ \ell$ and $ \eta _ { \ell ^ \prime }$ are the 
elastic scattering phase shifts of the collision system, respectively, 
$ C(\ell \ell^\prime L ; 0 0 ) $ the Clebsch-Gordan coefficients
\cite{Rose_1957}, $\theta$  the center of mass scattering angle,
and $ P_L (\cos\theta) $ the Legendre polynomial of the order $L$.
The angle integrated cross section $\sigma_\ell$ 
for partial wave $\ell$ is given by
\begin{equation}
\label{eq:PartialCrossSection}
\sigma_\ell = \frac{4\pi}{k^2}(2\ell+1)\sin^2\eta_\ell .
\end{equation}
When we have an isolated resonance at $ \ell = \ell_\mathrm{r} $,
we can represent the phase shifts $ \eta _ \ell $ in terms 
of the resonance phase shift $\Delta _ {\ell _ \mathrm{r}}$ and 
the background phase shift $ \delta _ \ell$
as follows.
\begin{equation}
\label{phaseshift}
\eta_\ell =  \delta_{\ell_\mathrm{r}} + \Delta_{\ell_\mathrm{r}} ~~ \mathrm{for} ~~ \ell = \ell_\mathrm{r}, ~~~ 
\eta_\ell = \delta_\ell ~~ \mathrm{for~else} .
\end{equation}
For a resonant partial wave $\ell = \ell_\mathrm{r} $, we have 
$\eta_\ell = \eta_{\ell_\mathrm{r}} = \delta_{\ell_\mathrm{r}} + \Delta_{\ell_\mathrm{r}} $ 
in eq.(\ref{eq:PartialCrossSection}). 
By introducing a reduced scattering energy
$\epsilon = -\cot\Delta_{\ell_\mathrm{r}} $ and 
a profile parameter
$q = -\cot\delta_{\ell_\mathrm{r}}$,
we obtain $\sigma_{\ell_\mathrm{r}}$ in the form of Fano profile formula\cite{Iizawa_2021}.
\begin{equation}
\label{eq:FanoProfileForPartialCrossSection}
\sigma_{\ell_\mathrm{r}} = \frac{4\pi}{k^2}(2\ell_\mathrm{r}+1)\frac{1}{(1+q^2)}\frac{(q+\epsilon)^2}{(1+\epsilon^2)} .
\end{equation}
The reduced energy $\epsilon$ is
related to the center-of-mass scattering energy $E$, 
resonance point energy $E_\mathrm{r}$, and 
the resonance width $\Gamma$ as $\epsilon = (E-E_\mathrm{r})/(\Gamma/2)$.

In the case of angle-differential cross-sections, 
the resonance profile is modified by the interference from
the partial waves with $ \ell \neq \ell_\mathrm{r} $.
Following the paper by Koike\cite{Koike_1977},
we can derive a resonance profile formula for  
${\mathrm{d}\sigma}/{\mathrm{d}\Omega}$
by introducing an angle-dependent complex-number parameter $w_{\ell_\mathrm{r}}$, which gives the characteristics of the resonance profile.
We take $ \mathrm{i} = \sqrt{-1}$. Then we have
\begin{eqnarray}
\label{eq:wparameter}
\noindent
w_{\ell_\mathrm{r}}(\theta) = \exp({2\mathrm{i}\delta_{\ell_\mathrm{r}}})
\sum_{L = 0}^{\infty}
\{(2\ell_\mathrm{r}+1)^2
[C(\ell_\mathrm{r}\ell_\mathrm{r}L;00)]^2\nonumber\\
+ 
\sum^\infty_{\ell=0,\ell \neq \ell_\mathrm{r}}(2\ell+1)(2\ell_\mathrm{r}+1)
[C(\ell\ell_\mathrm{r}L;00)]^2\nonumber \\
\times
[1-\exp(-2\mathrm{i}\delta_\ell)])\}
P_L(\cos\theta).
\end{eqnarray}
The cross-section ${\mathrm{d}\sigma}/{\mathrm{d}\Omega}$ can be given 
by a half of the real part of
\begin{equation}
\label{eq:Flr}
F_{\ell_\mathrm{r}} \equiv - \frac{1}{k^2}w_{\ell_\mathrm{r}}\exp(2\mathrm{i}\Delta_{\ell_\mathrm{r}}) + z
= - \frac{1}{k^2}w_{\ell_\mathrm{r}}\frac{\epsilon -\mathrm{i}}
{\epsilon + \mathrm{i}} + z ,   
\end{equation}
where $z$ is the background part cross-section.
The explicit expression of $z$ is given by\cite{Koike_1977}
\begin{equation}
\label{eq:z}
    z = \sum_{L=0}^{\infty}Z_LP_L(\cos(\theta)) ,  
\end{equation}
with
\begin{eqnarray}
\label{eq:Z_L}
\noindent
    Z_L = (2\ell_\mathrm{r} + 1)^2
    [C(\ell_\mathrm{r} \ell_\mathrm{r} L ; 0 0 ) ]^2
    ~~~~~~~~~~~~~~~~~~~~~~~~~~\nonumber \\
    +~ 2\sum_{\ell=0,\ell \neq \ell_\mathrm{r}}^{\infty}
    (2\ell + 1 )(2\ell_\mathrm{r} + 1)
    [C(\ell \ell_\mathrm{r} L ; 0 0 ) ]^2\sin^2\delta_\ell\nonumber \\
    +~ 2\sum_{\ell=0,\ell \neq \ell_\mathrm{r}}^{\infty}
    \sum_{\ell^\prime=0,\ell^\prime \neq \ell_\mathrm{r}}^{\infty}
    (2\ell + 1)(2\ell^\prime + 1)[C(\ell \ell^\prime L ; 0 0) ]^2\nonumber \\
    \times \sin\delta_\ell \sin\delta_{\ell^\prime} \cos (\delta_\ell - \delta_{\ell^\prime})  . 
\end{eqnarray}

Introducing the following quantity
\begin{equation}
\label{eq:qparameter}
q(\theta) \equiv -\cot\left(\frac{1}{2}
\mathrm{arg}(w_{\ell_\mathrm{r}}(\theta))\right),
\end{equation}
we have 
\begin{equation}
\label{eq:FanoProfile}
\frac{\mathrm{d}\sigma}{\mathrm{d}\Omega}
= \frac{1}{k^2}\left|w_{\ell_\mathrm{r}}\right|
\frac{1}{1+q^2}\frac{(q+\epsilon)^2}{1+\epsilon^2}
+ \frac{1}{2}(z-\left|w_{\ell_\mathrm{r}}\right|),
\end{equation}
where, $\mathrm{arg}(w_{\ell_\mathrm{r}}(\theta))$ in eq.(\ref{eq:qparameter})
gives the  
argument of the complex-number $w_{\ell_\mathrm{r}}(\theta)$.
The angle-dependent $q=q(\theta)$ reduces to the ordinary $q$-parameter as introduced in eq.(\ref{eq:FanoProfileForPartialCrossSection}) when all the phase shifts of non-resonant partial waves are vanishing. The $q=q(\theta)$ defined in eq.(\ref{eq:qparameter}) is a natural extension of the ordinary angle-independent $q$-parameter.

We can point out here that the resonance profile in the 
resonant elastic differential scattering cross section
can be represented by the Fano profile formula\cite{Fano_1961} 
with a scattering angle dependent parameter $q = q(\theta)$ reduced from the 
complex-number parameter $w_{\ell_\mathrm{r}}(\theta)$
as in eq.(\ref{eq:qparameter}). 
The absolute value of $w_{\ell_\mathrm{r}}$ is connected to the height of the resonance profile, and $ \mathrm{arg} (w_{\ell_\mathrm{r}}) $ gives the asymmetry of the resonance profile. Thus, the parameter $w_{\ell_\mathrm{r}}$
completely characterizes the angle dependence of the resonance profile.
Furthermore, as found in eq.(\ref{eq:wparameter}), 
we may note that $w_{\ell_\mathrm{r}}$ is effectively given by 
a polynomial of $\cos\theta$;  $w_{\ell_\mathrm{r}}$ is 
analytic with respect to $\cos \theta $ or $\theta$.
\section{\label{sec:H+Kr}Resonances in $ \mathrm{H ~ + ~ Kr}$ collisions}
Toennies et al\cite{Toennies_1979} studied the molecular beam scattering of hydrogen atoms/molecules from rare gas atoms,
which helped in modeling the inter-atomic interaction potentials. Extensive comparison with spherically symmetric model potentials has revealed that all the observed resonances can be characterized as orbiting resonances. 
The resonance energy positions are highly sensitive to the potential forms. 
They proposed the inter-atomic interaction potentials $V (r) $ 
as a function of the inter-atomic distance $ r $ of the following form.
\begin{eqnarray}
 \label{eq:potential}
 V (r) = \left\{
 \begin{array}{cc}
        D_e[(\frac{r_\mathrm{m}}{r})^{12} - 2(\frac{r_\mathrm{m}}{r})^6 ] ~~\mathrm{for}~ r < r_\mathrm{s} \\
        \\
        - \frac{C_6}{r^6} - \frac{C_8}{r^8} - \frac{C_{10}}{r^{10}}  ~~~~~~~\mathrm{for}~ r \geq r_\mathrm{s} 
 \end{array}
 \right\},
 \end{eqnarray}
where $D_e$ and $r_\mathrm{m}$ are the well depth and the equilibrium distance 
of the Lennard-Jones potential, respectively.
The parameters $C_6$, $C_8$, and $C_{10}$ are of the dispersion forces
at $r \geq r_\mathrm{s}$. 
In Table \ref{Potentials}, the potential parameters of 
$ \mathrm{H ~ + ~ Kr}$ collision system are tabulated. 
We use this set of parameters for scattering calculations in the present paper.

\begin{figure}
    \centering
    \includegraphics[width=7.5cm]{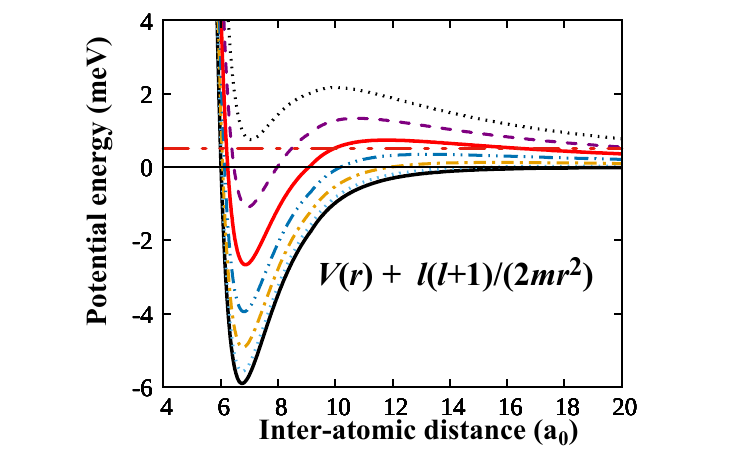}
    \caption{Effective inter-atomic interaction potential 
    $ V(r) + {\ell(\ell+1)}/{(2mr^2)} $ for $\ell = 0 ~ \mathrm{to} ~ 6$
    are shown in order of increasing potential energy.
    Solid(black online) curve at the bottom is of $ V(r)$ given in eq.(\ref{eq:potential}). Solid(red online) curve at the fifth from the bottom 
    is of the $\ell = 4$ partial wave, which undergoes the shape resonance.
    Horizontal double dot dashed(red online) line 
    at $0.512~ \mathrm{meV} = 4.13~ \mathrm{cm}^{-1}$ indicates the
    resonance point energy of $\ell = 4$ resonant partial wave. 
    Potential energy is given in units of meV. Inter-atomic distance is given in   
    units of $\mathrm{a_0}$(Bohr radius).
    } 
    \label{fig:H+Kr_Potential}
\end{figure}

In Fig.\ref{fig:H+Kr_Potential}, we show the effective inter-atomic interaction potential of the partial waves 
$V(r) + \ell (\ell + 1)/(2mr^2)$ for $\ell = 0 ~ \mathrm{to} ~ 6$, where $m$ is the reduced mass of the collision system. As discussed later in this section, the 
$\ell = 4 $ partial wave undergoes resonance at 
$ E = 0.512 ~\mathrm{meV} = 4.13 ~ \mathrm{cm}^{-1} $. The resonance point energy is indicated by thick (red online) double dot dashed horizontal line. At $\ell = 4$, the effective potential makes a slight hump around $ r = 12 ~\mathrm{a}_0 $ (Bohr radius). The $ \ell = 4$ partial wave is tunneling into the potential well to undergo the resonance.  
\begin{table}
\caption{\label{Potentials} The inter-atomic interaction potential 
of $ \mathrm{H ~ + ~ Kr}$ collision system\cite{Toennies_1979}.
For notations $r$, $D_e$, $r_\mathrm{m}$, $C_6$, $C_8$, $C_{10}$, and $r_\mathrm{s}$, see 
eqs.(\ref{eq:potential}) and their followings.
} 
%
%
\begin{tabular}{c|c|c}
%
\hline\hline
\textrm{Region}&
\textrm{Model}&
\textrm{Parameters}\\
\colrule
%
%
%
$r < r_\mathrm{s}$ & Lennard-Jones & $D_e = 5.90$ meV \\
 & & $ r_\mathrm{m} = 3.57 $ \AA \\
$r \geq r_\mathrm{s}$ & $-\frac{C_6}{r^6}-\frac{C_8}{r^8}-\frac{C_{10}}{r^{10}} $ &$C_6 = 17.1 \times 10^3$ meV\AA$^6$ \\
  & & $C_8 = 96.0 \times 10^3 $  meV\AA$^8$ \\
($r_\mathrm{s}=4.83$ \AA ) & & $C_{10} = 695.2 \times 10^3 $  meV\AA$^{10}$ \\
\hline\hline
\end{tabular}
\end{table}

To obtain the elastic scattering phaseshifts and cross sections,
we consider the following Schr\"{o}dinger equations for the $ \ell $-partial wavefunction $\lambda_{\ell}(r)$. 
We take $U (r) = 2mV(r)$.
We use atomic units $ e = \hbar = m_\mathrm{e} = 1$ throughout the paper unless otherwise stated.
We have 
\begin{equation}
\label{eq:Schroedinger} 
\frac{\mathrm{d}^2\lambda_\ell(r)}{\mathrm{d}r^2} 
+ \left\{k^2 - \frac{\ell(\ell+1)}{r^2}\right\}
\lambda_\ell(r)
= U(r)\lambda_\ell(r) 
\end{equation}
with the following asymptotic form.
\begin{equation}
\label{eq:asymptoticform}
\lambda_\ell(r)_{r \rightarrow \infty} \rightarrow A_\ell \sin(kr-\frac{\ell}{2}\pi+\eta_\ell) .
\end{equation}
By employing the variable phase approach(VPA)\cite{PALOV2021107895}, 
we can directly solve $\eta_\ell$ as a function of $r$. 
We have 
\begin{equation}
\label{eq:Phase Equation1}
\frac{\mathrm{d}\eta_\ell(r)}{\mathrm{d}r} = -\frac{1}{k}U(r)G_{\ell}^2(\eta_\ell(r), kr) 
\end{equation}
with
\begin{equation}
\label{eq:Phase Equation2}
G_\ell(\eta_\ell(r), kr) \equiv \cos(\eta_\ell(r)) \hat{j}_\ell(kr) - \sin(\eta_\ell(r)) \hat{n}_\ell(kr) ,
\end{equation}
where $ \hat{j}_\ell(kr)$ and $ \hat{n}_\ell(kr)$ are, respectively,
Ricatti-Bessel function of the first and second kind.
We solve eq.(\ref{eq:Phase Equation1}) with the boundary condition 
$\eta_\ell (0) = 0$; the scattering phaseshift $\eta_\ell$ is given by 
$\eta_\ell(r)$ at $r \rightarrow \infty$.
We employed a program VPA by Palov and Balint-Kurti\cite{PALOV2021107895}
in CPC program library for numerical calculation.
\begin{figure}
    \centering
    \includegraphics[width=7.5cm]{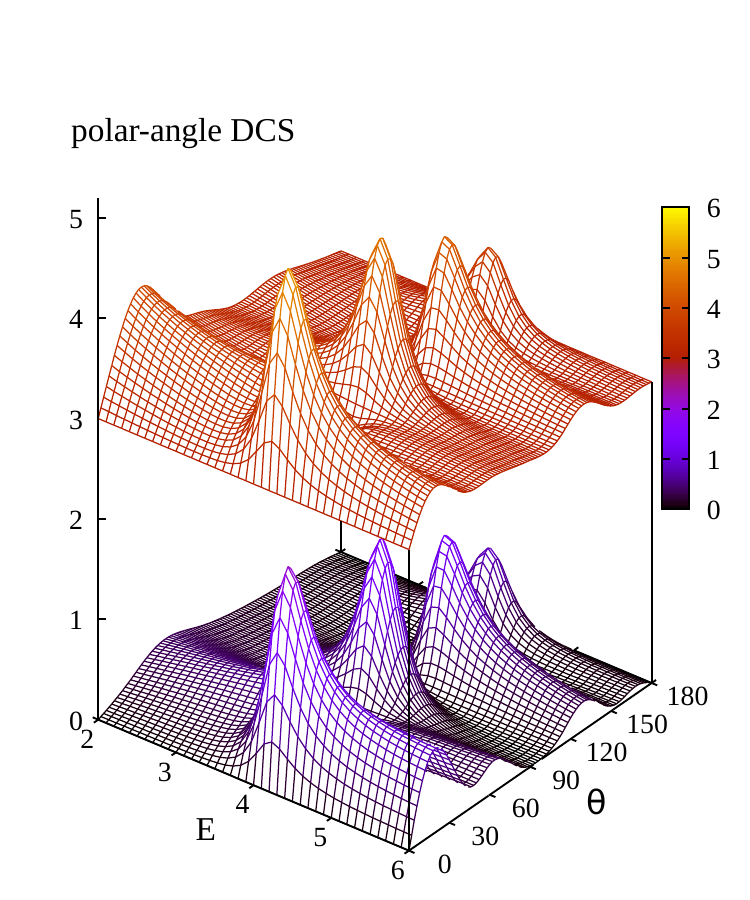}
    \caption{ 
    The energy dependence of the polar-angle DCS 
    $ \mathrm{d}\sigma/\mathrm{d}\theta$ defined in eq.(\ref{eq: Polar-angle DCS}).
    The lower entry: Blatt and Biedenharn's formula\cite{Blatt_1952} given in eq.(\ref{DifferentialCrossSection}).
    The upper entry: synthesized cross-section formula given in eq.(\ref{eq:FanoProfile}). 
    The values have been raised up by $ 3 \times 10^3 {\mathrm{a}_0}^2$ to give a separate view of the cross sections.
    On the horizontal plane, E: center of mass collision energy in units of $ \mathrm{cm}^{-1} $, and
    $\theta$: center of mass scattering angle in units of degree. Vertical axis: polar-angle DCS in units of $ 1 \times 10^3 {\mathrm{a}_0}^2 $ 
    } 
    \label{fig:H+Kr_angledcs}
\end{figure}
To investigate the resonance feature of $\ell = \ell_\mathrm{r} = 4$ wave, we calculate 
$\eta_\ell$ for $\ell = 0 ~\mathrm{to}~ 6$ and for the range of $E = 2.0 ~\mathrm{to}~ 6.0 ~\mathrm{cm}^{-1}$.
And, further, we confirmed that $\eta_\ell$ is negligibly small for $\ell > 6 $ in the
range of $E$ considered. Using the result of calculation, we evaluated the angle differential cross section 
$ \mathrm{d}\sigma/\mathrm{d}\Omega$ for the range of polar angle $ \theta = 0^\circ ~\mathrm{to}~ 180^\circ$.
Also, because we are considering the axial symmetric case, we evaluated the 
polar-angle differential cross section (polar-angle DCS) $\mathrm{d}\sigma/\mathrm{d}\theta$, which is given by
\begin{equation}
\label{eq: Polar-angle DCS}
\frac{\mathrm{d}\sigma}{\mathrm{d}\theta} = 2\pi\sin \theta \cdot (\frac{\mathrm{d}\sigma}{\mathrm{d}\Omega}).
\end{equation}

In Fig.\ref{fig:H+Kr_angledcs}, we show the energy and angle dependence of $ (\mathrm{d}\sigma/\mathrm{d}\theta)$
for the range of $E = 2.0 ~\mathrm{to}~ 6.0 ~\mathrm{cm}^{-1}$ and $\theta = 0^\circ ~\mathrm{to}~ 180^\circ$.
By plotting the polar-angle DCS, we can give a better appearance over the whole range of 
$ \theta $ except the forward and backward extremes.
We can clearly observe the presence of resonance around $E = 4.13 ~\mathrm{cm}^{-1}$.
The asymmetry of the resonance profile varies with the change of $\theta$ due to the 
angle-dependent strength of the interference from the non-resonant partial waves, which will be discussed in detail later in this section.
The lower entry of the figure gives the cross-section
by Blatt and Biedenharn's formula\cite{Blatt_1952} given in eq.(\ref{DifferentialCrossSection}).
The upper entry represents the synthesized cross section by the formula
in eq.(\ref{eq:FanoProfile}). The values are raised up by $ 3000 ~ \mathrm{a}_0^2$
to give a separate view of the cross sections. We can point out 
that eq.(\ref{eq:FanoProfile}) simulates well the rigorous cross sections by Blatt and Biedenharn[\cite{Blatt_1952}.

\begin{figure}
    \centering
    \includegraphics[width=7.5cm]{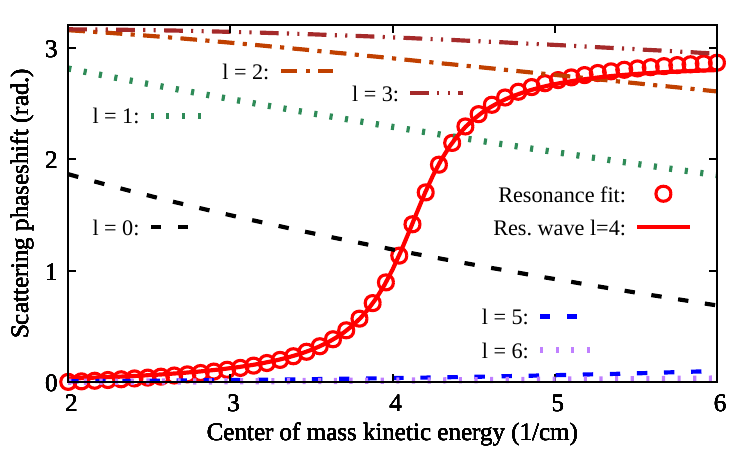}
    \caption{Partial wave phaseshifts $ \eta_\ell$ for $ \ell = 0 ~\mathrm{to}~ 6$ and resonance phaseshift analysis for $ \ell _\mathrm{r} = 4 $.
    Dashed (black online) curve: $ \ell = 0 $, 
    Dotted (sea-green online) curve: $ \ell = 1 $,
    Dot dashed (dark-orange online) curve: $ \ell = 2 $,
    Double dot dashed (brown online) curve: $ \ell = 3 $,
    Solid (red online) curve: $ \ell = 4 $,
    Dashed (blue online) curve: $ \ell = 5 $,
    Dotted (purple online) curve: $ \ell = 6 $.
    Open (red online) circles show the fitting of the resonance formula
    $ \eta _ {\ell_\mathrm{r}} = \delta _ {\ell_\mathrm{r}} + \Delta _ {\ell_\mathrm{r}} = \delta _ {\ell_\mathrm{r}} - \cot^{-1}((E - E_\mathrm{r})/(\Gamma/2)) $ 
    to the VPA-calculation for $ \ell = 4 $ partial wave.
    } 
    \label{fig:H+Kr_psplot}
\end{figure}

\begin{figure}
    \centering
    \includegraphics[width=7.5cm]{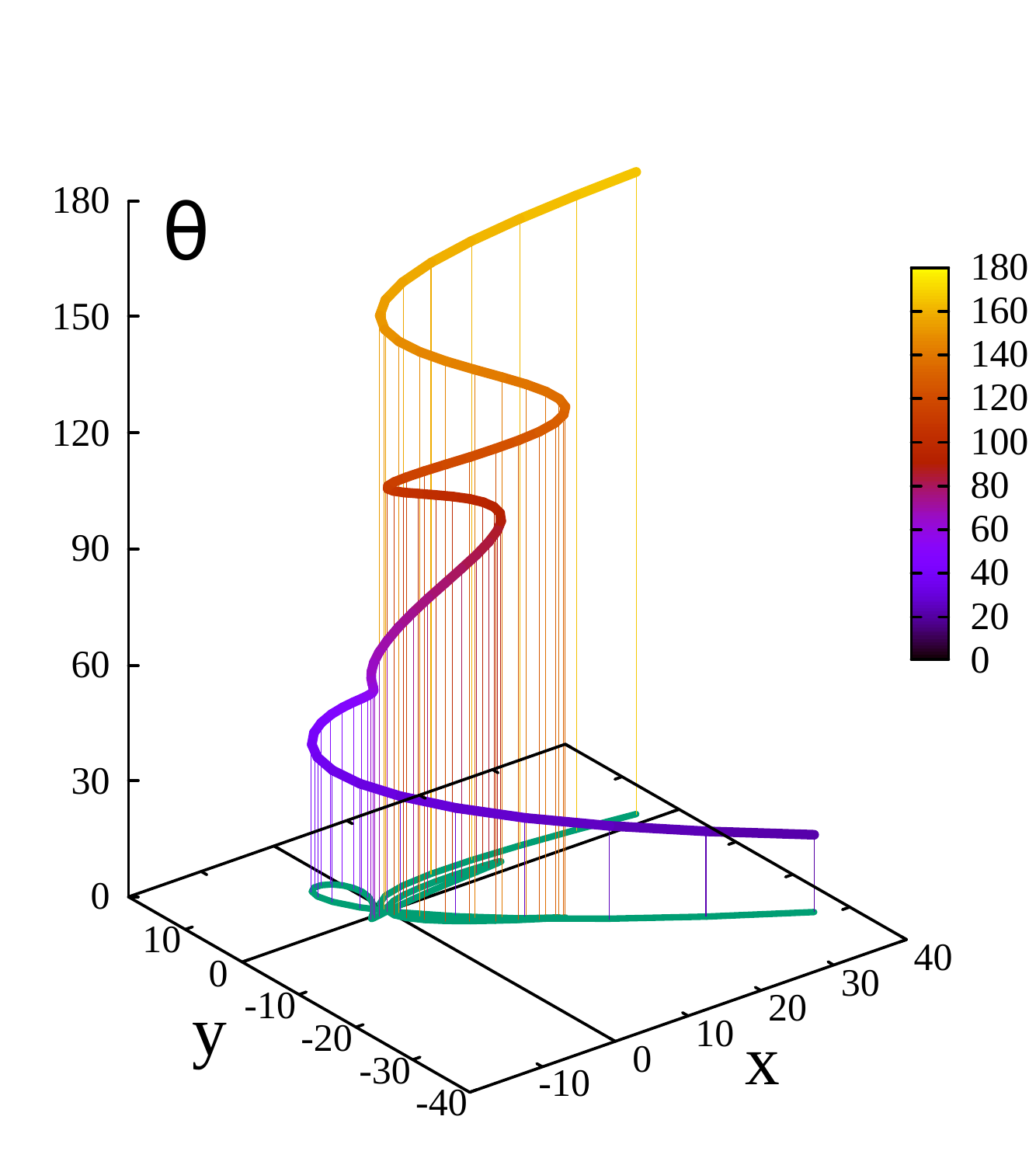}
    \caption{Three dimensional scattering angle dependent evolution of the complex-number parameter $w_{\ell_\mathrm{r}}(\theta)$ 
    (See eq.(\ref{eq:wparameter})). 
    The set of horizontal axes $x$ and $y$ represents the Gauss plane $ x + \mathrm{i}y$. 
    The vertical axis gives the scattering angle $\theta$ in units of degree. 
    Thick solid curve with (color online) gradation:  $w_{\ell_\mathrm{r}}(\theta)$.
    The gradation(color online) illustrates the evolution of $\theta $.
    Solid (green online) curve: Projection of $w_{\ell_\mathrm{r}}(\theta)$ onto the Gauss plane.  
    }
    \label{fig:H+Kr_wprmsteric}
\end{figure}

To further investigate the effect of angle-dependent interference on the resonance profile,
we analyze the behavior of the partial wave phaseshift  $\eta_\ell $ in detail.
In Fig.\ref{fig:H+Kr_psplot}, we show $ \eta _ \ell $ for $\ell = 0 ~\mathrm{to}~ 6$.
We can observe a sharp resonance structure 
at $\ell = \ell_{\mathrm{r}} = 4$,
which was pointed out as an orbiting resonance by Toennies\cite{Toennies_1979}.
We performed a resonance phaseshift analysis to the $ \ell _ \mathrm{r} $ partial wave.
We fitted the resonance formula $ \eta _ {\ell_\mathrm{r}} = \delta _ {\ell_\mathrm{r}} + \Delta _ {\ell_\mathrm{r}}= \delta _ {\ell_\mathrm{r}} - \cot^{-1}((E - E_\mathrm{r})/(\Gamma/2)) $ 
given in eq.(\ref{phaseshift}) to the VPA-calculation.
We found that the set of $ \delta _ {\ell_\mathrm{r}} = - 0.128~ \mathrm{radian}$,
$ E_\mathrm{r} = 4.13~ \mathrm{cm}^{-1}$, and $ \Gamma = 0.56~ \mathrm{cm}^{-1} $
gives a good fit to the calculated $ \eta_{\ell_\mathrm{r}}$,
which is given by a solid (red online) curve in Fig.\ref{fig:H+Kr_psplot}.
The fitted resonance curve is plotted by open (red online) circles in the figure.
At the resonance point $ E_\mathrm{r} = 4.13~ \mathrm{cm}^{-1}$, we evaluated the background phaseshifts
$ \delta _ \ell $ for $ \ell = 0 ~\mathrm{to}~ 6 $. They are listed in Table \ref{tab:Phase Shift and Resonance Parameter} with the obtained resonance parameters.
%
\begin{table}
\caption{\label{tab:Phase Shift and Resonance Parameter} 
        Background phaseshifts $\delta_\ell$ for $\ell = 0 ~ \mathrm{to} ~ 6$ at $ E = E _ \mathrm{r} = 4.13 ~\mathrm{cm}^{-1}$,
and the resonance parameters $E_\mathrm{r}$ and $\Gamma$ for $\ell_\mathrm{r} = 4 $ resonant partial wave. 
}
\begin{tabular}{c|ccc}
\hline\hline
$~~~\ell~~~$ & $\delta_\ell$ ( radian )  & $E_\mathrm{r}$ ($\mathrm{cm}^{-1}$) & $\Gamma$ ($\mathrm{cm}^{-1}$) \\ 
\hline
$0$  & ~1.153 &       &      \\
$1$  & ~2.258 &       &      \\
$2$  & ~2.884 &       &      \\
$3$  & ~3.084 &       &      \\
$4$  & -0.128 & 4.13  & 0.56 \\
$5$  & ~0.040 &       &      \\
$6$  & ~0.015 &       &      \\
\hline\hline
\end{tabular}
\end{table}
Using the values given in Table \ref{tab:Phase Shift and Resonance Parameter},
we evaluate $w_{\ell_{\mathrm{r}}}(\theta)$ in eq.(\ref{eq:wparameter}) and 
then $q(\theta)$ in eq.(\ref{eq:qparameter}). 
We used a program ANGMOM (catalogue number: ACYK) in
CPC Program Library for Clebsch-Gordan coefficients\cite{Rao1978}.
In Fig.\ref{fig:H+Kr_wprmsteric}, we show the
steric view of the $\theta$ evolution of  
$w_{\ell_\mathrm{r}}(\theta) $
for $\ell_\mathrm{r} = 4$ resonance at $E_\mathrm{r} = 4.13 ~\mathrm{cm}^{-1}$
on a Gauss plane by solid curve with (color online) gradation.
The set of horizontal axes $x$ and $y$ represents the Gauss plane $ x + \mathrm{i}y$.
The vertical axis gives the scattering angle $\theta$ in units of degree.
We also give the projection of $w_{\ell_\mathrm{r}}(\theta) $ onto the Gauss plane
by solid (green online) curve.
We find that $w_{\ell_\mathrm{r}}(\theta) $ is continuous and smooth
as a function of $\theta$.
We can thus visualize the interference feature of the shape and magnitude for the resonance profile.

\begin{figure}
     \centering
     \includegraphics[width=7.5cm]{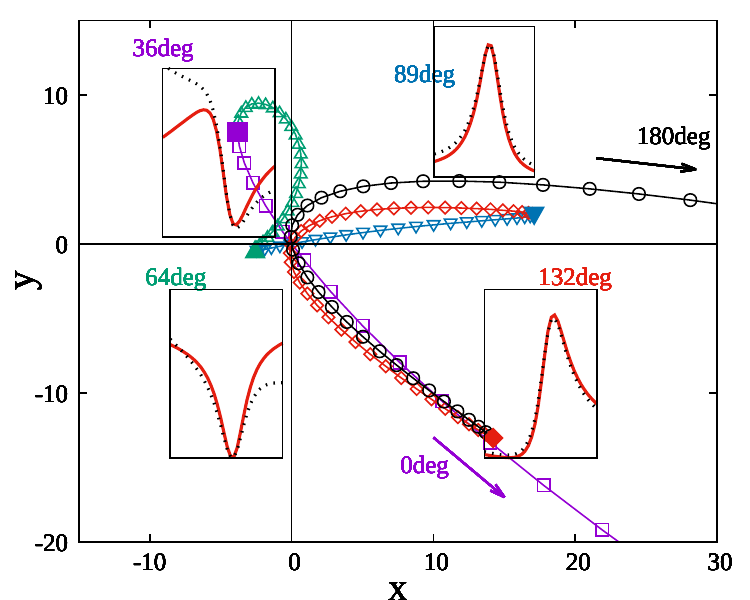}
     \caption{The trace of $w_{\ell_{\mathrm{r}}}(\theta)$ for $\theta = 20^\circ ~\mathrm{to}~ 160^\circ$ on the Gauss plane $ x + \mathrm{i}y$.
     Open square (blue online): $\theta \leq 36^\circ$, 
     Open triangle (green online): $36^\circ < \theta \leq 64^\circ$,
     Open inverse triangle (light blue online):  $64^\circ < \theta \leq 89^\circ$,
     Open rhombus (red online):  $89^\circ < \theta \leq 132^\circ$, 
     and, 
     Open circle (black online):  $132^\circ < \theta$. 
     Marks of the representative scattering angles are 
     solid (blue online) square on $ 36^\circ$,
     solid (green online) triangle on $ 64^\circ$,
     solid (light blue online) inverted triangle on $ 89^\circ$,
     and
     solid (red online) rhombus for $132^\circ$.
     The insets show the Fano resonance profiles at those angles.
     Solid curve: eq.(\ref{eq:FanoProfile}), and
     Dotted curve: eq.(\ref{DifferentialCrossSection}).
     }
     \label{fig:H+Kr_wprmwithprofile}
 \end{figure}

To obtain a deeper insight into the resonance profile at moderate scattering angles,
we evaluate the resonance profiles at several representative $ \theta $. 
In Fig.\ref{fig:H+Kr_wprmwithprofile}, we show the trace of $w_{\ell_\mathrm{r}}(\theta)$ 
for $\theta = 23^\circ ~ \mathrm{to} ~ 164^\circ $.
Because the resonance is on $\ell_\mathrm{r} = 4$ and 
$|w_{\ell_\mathrm{r}}(\theta)|$ is its height, $w_{\ell_\mathrm{r}}(\theta)$ vanishes 
$ 4 $ times through the whole range of $\theta$, as discussed later in detail in the next section.
In this figure, to visualize the angle evolution of  $w_{\ell_\mathrm{r}}(\theta)$ clearly,
we divided the range of angles into several parts and illustrated the curve with different symbols and colors. 
For several representative scattering angles,
which are marked on the $w_{\ell_\mathrm{r}}(\theta)$ curve by 
solid (blue online) square for $ 36^\circ $,
solid (green online) triangle for $ 64^\circ$,
solid (light blue online) inverted triangle for $ 89^\circ$,
and
solid (red online) rhombus for $132^\circ$,
the corresponding normalized Fano profiles are given in the insets
by (red online) solid curves (eq.(\ref{eq:FanoProfile})). 
For comparison, we also give the correspondent angle differential cross-section
(eq.(\ref{DifferentialCrossSection})) by (blue online) dots.
The range of $E$ is from $3.0$ to $5.0 ~\mathrm{cm}^{-1}$.

\section{\label{sec:Discussion}Discussion}

Fano resonance is known as one of the ubiquitous phenomena that are observed in isolated resonators that are embedded in broadband oscillations regardless of quantum or classical systems\cite{Iizawa_2021}.
The Fano resonance in cold atomic collisions gives deep insight into the relation of quantum interference with inter-atomic interaction potential.
Resonances in the elastic scattering of quantum mechanical particles can usually be observed when the de Broglie wavelength is  
comparable to the range of interaction potential. For cold atomic collisions, their de Broglie wavelengths are long enough to satisfy  
such a condition. Because the resonance behavior is highly sensitive to the inter-atomic interaction potentials,
the analysis of resonance profiles can provide a demanding test for potential evaluations.

In the present paper, we have pointed out that the angle-dependent evolution of the resonance profiles for cold atomic and molecular collisions can be
characterized by a complex-valued parameter $ w_{\ell_\mathrm{r}} $ given in eq.(\ref{eq:wparameter}),
following the discussion by Koike \cite{Koike_1977} on the resonance in slow electron scattering from atoms.
The height of the resonance is given by the absolute value of $ w_{\ell_\mathrm{r}}(\theta) $, as seen in eq.(\ref{eq:FanoProfile}). 
We also find from eqs.(\ref{eq:qparameter} and \ref{eq:FanoProfile}) that the angle-dependent asymmetry of the resonance profile 
can be characterized by the argument of the complex-number $w_{\ell_\mathrm{r}} (\theta) $; Fano's asymmetry parameter $q$ 
is expressed in terms of $\mathrm{arg}  w_{\ell_\mathrm{r}} (\theta) $ as in eq(\ref{eq:qparameter}). 
The angle-dependent asymmetry of the resonance can be described by a single Fano profile function 
$ (q + \epsilon)^2/(1 + \epsilon^2)$ with an angle-dependent $q$-parameter $ q(\theta) $.  
\begin{figure}
     \centering
     \includegraphics[width=7.5cm]{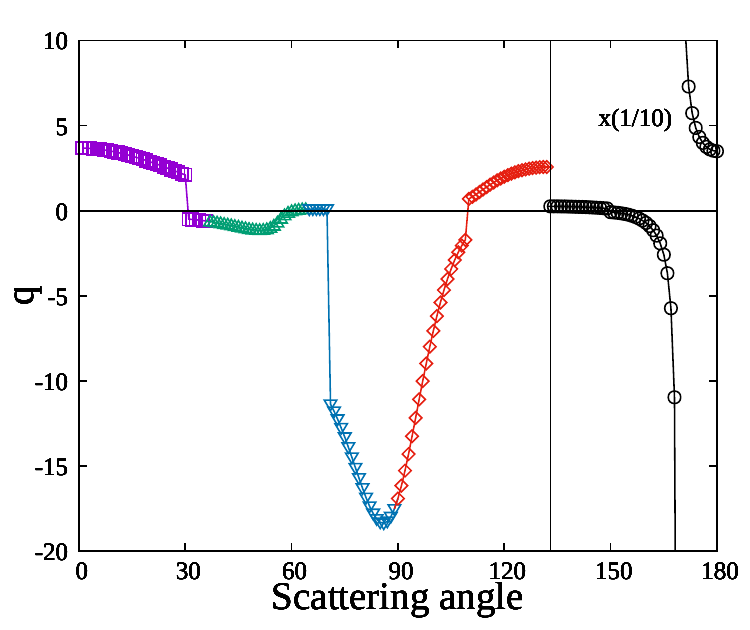}
     \caption{Angle dependent Fano's q-parameter $ q (\theta) $ for $\ell_\mathrm{r} = 4 $
     shape resonance of $ \mathrm{H} + \mathrm{Kr} $ elastic scattering.
     Horizontal axis: Scattering angle $\theta $ in units of $\mathrm{degree}$.
     Vertical axis: $q (\theta) $.
     For the symbols on the curve, see the caption in Fig.\ref{fig:H+Kr_wprmwithprofile}.
     }
     \label{fig:H+Kr_qprmongraph}
\end{figure}
We can calculate $ q(\theta)$ using eq.(\ref{eq:qparameter}).
In Fig.\ref{fig:H+Kr_qprmongraph}, we plot $ q(\theta)$ for $\theta = 0^\circ ~\mathrm{to}~ 180^\circ$.
As found in this plot, $ q(\theta)$ becomes discontinuous as a function of $ \theta $ 
at several values of $ \theta $. and diverges at $ \theta = 169^\circ $.
These discontinuities in the $q$-parameter as a function of $\theta$ are related to the zero points of the $\theta$-dependent partial wave functions $P_{\ell_\mathrm{r}} (\cos(\theta))$ of the resonant partial waves.
For example, for $\ell_\mathrm{r} =4, w_{\ell_\mathrm{r}}(\theta) = 0$ at $\theta = 30.56\,\mathrm{degree}$ (see Fig.\ref{fig:H+Kr_wprmwithprofile}), 
and as $\theta$ increases and crosses the origin of the Gaussian plane, 
$w_{\ell_\mathrm{r}}(\theta)$ moves from quadrant 4, where $q>0$, to quadrant 2, where $q<0$.
Thus, the $q$-parameter changes discontinuously at the point $w_{\ell_\mathrm{r}}(\theta) = 0$.
We have to point out that this feature could cause difficulties in the resonance profile analysis of experimental spectra.
Because the experiments must contain the data from a certain range of $ \theta $ 
due to the resolution of the apparatus, it might sometimes be inappropriate to assign
a single value of $q$ for an experimental resonance profile.
In contrast, as confirmed in eq.(\ref{eq:wparameter}), Figs. \ref{fig:H+Kr_wprmsteric}, and \ref{fig:H+Kr_wprmwithprofile},
the parameter $w_{\ell_\mathrm{r}} (\theta)$ is continuous and smooth throughout the whole range of $\theta$,
which may ensure the stability of the spectral analysis.

We theoretically investigated the profile asymmetry of $ \ell_\mathrm{r} = 4 $ orbiting resonance for $ \mathrm{H} + \mathrm{Kr} $ 
elastic scattering around $ E_\mathrm{r} = 4.13 ~\mathrm{cm}^{-1} $.
As found in Figs. \ref{fig:H+Kr_wprmsteric} and \ref{fig:H+Kr_wprmwithprofile},  $ w_{\ell_\mathrm{r}}(\theta) $ makes five wings 
in the scattering angles reflected by the character of $\ell_\mathrm{r} = 4 $ resonance  wave, of which angular part is given by 
$ Y_{\ell,0}(\theta, \phi) = \frac{3}{2\sqrt{\pi}}P_4 (\cos \theta )$,
where $Y$ and $\phi$ are the spherical harmonics and the azimuth angle with respect to the incident beam direction, respectively.
As seen in Fig. \ref{fig:H+Kr_wprmwithprofile},
$ w_{\ell_\mathrm{r}}(\theta) $ becomes zero for $4$ times at the zero points of $ P_4(\cos \theta)$, 
$ \theta = 30.56^\circ, 70.12^\circ, 109.88^\circ, ~\mathrm{and}~149.44^\circ$, because the resonance height is zero when the amplitude of resonant partial wave is zero. 
At $ w_{\ell_\mathrm{r}}(\theta) = 0 $, the resonance profile vanishes and the parameter $q(\theta)$ experiences a discontinuous change (see Fig.\ref{fig:H+Kr_qprmongraph}) at its passage of zero point. For example, at $\theta = 31^\circ$, $ w_{\ell_\mathrm{r}}(\theta) $ passes through the coordinate origin from the 4th quadrant to the 2nd quadrant on the Gauss plane with the increase of $\theta$. The value of $q(\theta)$ jumps from $2.4$ to $-0.4$ discontinuously,  
whereas 
$ w_{\ell_\mathrm{r}}(\theta) $ is continuous and smooth at the passage.
Furthermore, we have to point out that $q(\theta)$ is divergent when $w_{\ell_\mathrm{r}} (\theta)$ is real and positive.
On the positive real axis of the Gauss plane, $\mathrm{arg} (w_{\ell_\mathrm{r}} (\theta))$ is $ 0 ~(\mathrm{or}~ 2\pi)$, and $q (\theta) \rightarrow \pm \infty $, accordingly.
For example, we observe the divergence of $q$ at $\theta = 169^\circ$ 
as seen in Fig. \ref{fig:H+Kr_qprmongraph}.
In fact, $w_{\ell_\mathrm{r}} (169^\circ) = 46.91 + 0.1875 i$ and
$w_{\ell_\mathrm{r}} (170^\circ) = 50.47 - 0.4131 i$.
The parameter $ w_{\ell_\mathrm{r}}(\theta) $ crosses the 
positive real axis in between, 
and, therefore, $q(\theta)$ diverges between $ \theta = 169^\circ ~\mathrm{and}~ 170^\circ$.
However, there are no physical significance for this divergence, because both $ q = \pm \infty $ indicate the same symmetric Lorentzian profile. 
In contrast, the parameter $w_{\ell_{\mathrm{r}}(\theta)}$ has no such singularity of the behavior along the evolution of $\theta$.
This suggests that the use of $ w_{\ell_\mathrm{r}}(\theta) $ is advantageous for the analysis of experimental resonance profiles compared to following the $\theta$ dependence of $q(\theta)$.

Paliwal et al \cite{Paliwal_2021} have carried out a challenging measurement on the Fano interference in angle-resolved elastic $\mathrm{He}^* + \mathrm{D}_2$ scattering for the $\ell_{\mathrm{r}} = 6$ partial wave at the resonance point $ E_\mathrm{r} = 4.8 ~ \mathrm{Kelvin}$. They found the angle-dependent evolution of the asymmetry in the resonance profile. 
They assigned it as the Fano resonance. They opened a wide area of research field in cold atomic resonance scattering.
They proposed, however, a profile formula that does not meet the conventional understandings for the value of $q$-parameter.
Fano's profile function $(q + \epsilon)^2/(1 + \epsilon^2)$ gives a window-type profile at $q = 0$, whereas their formula gives Lorentzian instead. 
It is, however, desirable to maintain the current convention on the values of $ q $ for the assignment of the shape of resonance. 
As shown in eq.(\ref{eq:FanoProfile}) of the present paper, the resonance profile can be represented rigorously by the conventional shape of Fano profile formula. 
It would be advantageous for further development of the resonance study in cold collisions to use a set of formulae that are analytic and maintain connectivity with the conventional theory of Fano resonance\cite{Fano_1961}. 
And, furthermore, we note that $q (\theta)$ of their resonance may change its sign for 6 times or more during the whole passage of $\theta$ from
$ 0^\circ $ to $ 180^\circ$, because they identified the resonance was of $ \ell_\mathrm{r} = 6$. 
The use of $w_{\ell_\mathrm{r}} (\theta)$ in the spectral analysis may give an advantage for overcoming the limited angle resolution of the experimental apparatus.

\section{\label{sec:Conclusion}Conclusion}
For angle-differential elastic processes in cold atomic collisions, the resonance line shape has been rigorously shown to be given by Fano's profile formula
$(q + \epsilon)^2/(1 + \epsilon^2)$ with an angle-dependent Fano's profile parameter $q = q(\theta) $. Explicit expression of $q(\theta) $
has been derived by introducing a complex number parameter $ w_{\ell_\mathrm{r}} $ which is a smooth function of $ \theta $ on the Gauss plane.
The resonance parameters $q(\theta)$ and $w_{\ell_\mathrm{r}}(\theta)$ were derived in a manner analogous to a previous study on slow electron-atom collisions, 
where the de Broglie wavelength of the collision system is similar to that of cold atomic and molecular collisions. 
Our paper presents new possibilities for our theoretical formalism, going beyond the previous electron-atom collision study to angle-dependent resonance profile analysis in the newly developing field of cold atomic and molecular collisions.

The Fano resonances in $ \mathrm{H} + \mathrm{Kr} $ differential elastic cold collisions have been theoretically studied using the potential given by Toennies et al\cite{Toennies_1979}. The resonance parameters $ q ( \theta ) $ and $ w_{\ell_\mathrm{r}} ( \theta ) $ are quite sensitive to the form of interaction potentials,
providing us with an opportunity for effective tests for the potential evaluation.    

\begin{acknowledgments}
For numerical calculations, the programs 
ANGMOM by K. Srinivasa Rao and K. Venkatesh\cite{Rao1978} and VPA by A. P. Palov and G. G. Balint-Kurti\cite{PALOV2021107895} in CPC Program
Library have been used.
One of the authors(FK) thanks to Prof. Y. Itoh of Josai University
for his useful discussions and suggestions.
\end{acknowledgments}
%

\begin{thebibliography}{22}%
\makeatletter
\providecommand \@ifxundefined [1]{%
 \@ifx{#1\undefined}
}%
\providecommand \@ifnum [1]{%
 \ifnum #1\expandafter \@firstoftwo
 \else \expandafter \@secondoftwo
 \fi
}%
\providecommand \@ifx [1]{%
 \ifx #1\expandafter \@firstoftwo
 \else \expandafter \@secondoftwo
 \fi
}%
\providecommand \natexlab [1]{#1}%
\providecommand \enquote  [1]{``#1''}%
\providecommand \bibnamefont  [1]{#1}%
\providecommand \bibfnamefont [1]{#1}%
\providecommand \citenamefont [1]{#1}%
\providecommand \href@noop [0]{\@secondoftwo}%
\providecommand \href [0]{\begingroup \@sanitize@url \@href}%
\providecommand \@href[1]{\@@startlink{#1}\@@href}%
\providecommand \@@href[1]{\endgroup#1\@@endlink}%
\providecommand \@sanitize@url [0]{\catcode `\\12\catcode `\$12\catcode `\&12\catcode `\#12\catcode `\^12\catcode `\_12\catcode `\%12\relax}%
\providecommand \@@startlink[1]{}%
\providecommand \@@endlink[0]{}%
\providecommand \url  [0]{\begingroup\@sanitize@url \@url }%
\providecommand \@url [1]{\endgroup\@href {#1}{\urlprefix }}%
\providecommand \urlprefix  [0]{URL }%
\providecommand \Eprint [0]{\href }%
\providecommand \doibase [0]{https://doi.org/}%
\providecommand \selectlanguage [0]{\@gobble}%
\providecommand \bibinfo  [0]{\@secondoftwo}%
\providecommand \bibfield  [0]{\@secondoftwo}%
\providecommand \translation [1]{[#1]}%
\providecommand \BibitemOpen [0]{}%
\providecommand \bibitemStop [0]{}%
\providecommand \bibitemNoStop [0]{.\EOS\space}%
\providecommand \EOS [0]{\spacefactor3000\relax}%
\providecommand \BibitemShut  [1]{\csname bibitem#1\endcsname}%
\let\auto@bib@innerbib\@empty
\bibitem [{\citenamefont {Fano}(1961)}]{Fano_1961}%
  \BibitemOpen
  \bibfield  {author} {\bibinfo {author} {\bibfnamefont {U.}~\bibnamefont {Fano}},\ }\bibfield  {title} {\bibinfo {title} {Effects of configuration interaction on intensities and phase shifts},\ }\href {https://doi.org/10.1103/PhysRev.124.1866} {\bibfield  {journal} {\bibinfo  {journal} {Phys. Rev.}\ }\textbf {\bibinfo {volume} {124}},\ \bibinfo {pages} {1866} (\bibinfo {year} {1961})}\BibitemShut {NoStop}%
\bibitem [{\citenamefont {Iizawa}\ \emph {et~al.}(2021)\citenamefont {Iizawa}, \citenamefont {Kosugi}, \citenamefont {Koike},\ and\ \citenamefont {Azuma}}]{Iizawa_2021}%
  \BibitemOpen
  \bibfield  {author} {\bibinfo {author} {\bibfnamefont {M.}~\bibnamefont {Iizawa}}, \bibinfo {author} {\bibfnamefont {S.}~\bibnamefont {Kosugi}}, \bibinfo {author} {\bibfnamefont {F.}~\bibnamefont {Koike}},\ and\ \bibinfo {author} {\bibfnamefont {Y.}~\bibnamefont {Azuma}},\ }\bibfield  {title} {\bibinfo {title} {The quantum and classical fano parameter q},\ }\href {https://doi.org/10.1088/1402-4896/abe580} {\bibfield  {journal} {\bibinfo  {journal} {Physica Scripta}\ }\textbf {\bibinfo {volume} {96}},\ \bibinfo {pages} {055401} (\bibinfo {year} {2021})}\BibitemShut {NoStop}%
\bibitem [{\citenamefont {Wigner}(1946{\natexlab{a}})}]{Wigner_PhysRev.70.15}%
  \BibitemOpen
  \bibfield  {author} {\bibinfo {author} {\bibfnamefont {E.~P.}\ \bibnamefont {Wigner}},\ }\bibfield  {title} {\bibinfo {title} {Resonance reactions and anomalous scattering},\ }\href {https://doi.org/10.1103/PhysRev.70.15} {\bibfield  {journal} {\bibinfo  {journal} {Phys. Rev.}\ }\textbf {\bibinfo {volume} {70}},\ \bibinfo {pages} {15} (\bibinfo {year} {1946}{\natexlab{a}})}\BibitemShut {NoStop}%
\bibitem [{\citenamefont {Wigner}(1946{\natexlab{b}})}]{Wigner_PhysRev.70.606}%
  \BibitemOpen
  \bibfield  {author} {\bibinfo {author} {\bibfnamefont {E.~P.}\ \bibnamefont {Wigner}},\ }\bibfield  {title} {\bibinfo {title} {Resonance reactions},\ }\href {https://doi.org/10.1103/PhysRev.70.606} {\bibfield  {journal} {\bibinfo  {journal} {Phys. Rev.}\ }\textbf {\bibinfo {volume} {70}},\ \bibinfo {pages} {606} (\bibinfo {year} {1946}{\natexlab{b}})}\BibitemShut {NoStop}%
\bibitem [{\citenamefont {Gibson}\ and\ \citenamefont {Dolder}(1969)}]{Gibson_1969}%
  \BibitemOpen
  \bibfield  {author} {\bibinfo {author} {\bibfnamefont {J.~R.}\ \bibnamefont {Gibson}}\ and\ \bibinfo {author} {\bibfnamefont {K.~T.}\ \bibnamefont {Dolder}},\ }\bibfield  {title} {\bibinfo {title} {Resonant differential scattering of electrons by helium},\ }\href {https://doi.org/10.1088/0022-3700/2/7/302} {\bibfield  {journal} {\bibinfo  {journal} {Journal of Physics B: Atomic and Molecular Physics}\ }\textbf {\bibinfo {volume} {2}},\ \bibinfo {pages} {741} (\bibinfo {year} {1969})}\BibitemShut {NoStop}%
\bibitem [{\citenamefont {Koike}(1977)}]{Koike_1977}%
  \BibitemOpen
  \bibfield  {author} {\bibinfo {author} {\bibfnamefont {F.}~\bibnamefont {Koike}},\ }\bibfield  {title} {\bibinfo {title} {New angle-dependent resonance profile parameter. i. electron-atom elastic scattering},\ }\href {https://doi.org/10.1088/0022-3700/10/14/021} {\bibfield  {journal} {\bibinfo  {journal} {Journal of Physics B: Atomic and Molecular Physics}\ }\textbf {\bibinfo {volume} {10}},\ \bibinfo {pages} {2883} (\bibinfo {year} {1977})}\BibitemShut {NoStop}%
\bibitem [{\citenamefont {Toennies}\ \emph {et~al.}(1979)\citenamefont {Toennies}, \citenamefont {Welz},\ and\ \citenamefont {Wolf}}]{Toennies_1979}%
  \BibitemOpen
  \bibfield  {author} {\bibinfo {author} {\bibfnamefont {J.~P.}\ \bibnamefont {Toennies}}, \bibinfo {author} {\bibfnamefont {W.}~\bibnamefont {Welz}},\ and\ \bibinfo {author} {\bibfnamefont {G.}~\bibnamefont {Wolf}},\ }\bibfield  {title} {\bibinfo {title} {Molecular beam scattering studies of orbiting resonances and the determination of van der waals potentials for h–ne, ar, kr, and xe and for h2–ar, kr, and xe},\ }\href {https://doi.org/10.1063/1.438414} {\bibfield  {journal} {\bibinfo  {journal} {The Journal of Chemical Physics}\ }\textbf {\bibinfo {volume} {71}},\ \bibinfo {pages} {614} (\bibinfo {year} {1979})}\BibitemShut {NoStop}%
\bibitem [{\citenamefont {Chilcott}\ \emph {et~al.}(2022)\citenamefont {Chilcott}, \citenamefont {Croft}, \citenamefont {Thomas},\ and\ \citenamefont {Kj\ae{}rgaard}}]{Chilcott_2022}%
  \BibitemOpen
  \bibfield  {author} {\bibinfo {author} {\bibfnamefont {M.}~\bibnamefont {Chilcott}}, \bibinfo {author} {\bibfnamefont {J.~F.~E.}\ \bibnamefont {Croft}}, \bibinfo {author} {\bibfnamefont {R.}~\bibnamefont {Thomas}},\ and\ \bibinfo {author} {\bibfnamefont {N.}~\bibnamefont {Kj\ae{}rgaard}},\ }\bibfield  {title} {\bibinfo {title} {Microscopy of an ultranarrow feshbach resonance using a laser-based atom collider: A quantum defect theory analysis},\ }\href {https://doi.org/10.1103/PhysRevA.106.023303} {\bibfield  {journal} {\bibinfo  {journal} {Phys. Rev. A}\ }\textbf {\bibinfo {volume} {106}},\ \bibinfo {pages} {023303} (\bibinfo {year} {2022})}\BibitemShut {NoStop}%
\bibitem [{\citenamefont {Li}\ \emph {et~al.}(2021)\citenamefont {Li}, \citenamefont {Hai}, \citenamefont {Lyu}, \citenamefont {Wang},\ and\ \citenamefont {Cong}}]{Li_2021}%
  \BibitemOpen
  \bibfield  {author} {\bibinfo {author} {\bibfnamefont {L.-H.}\ \bibnamefont {Li}}, \bibinfo {author} {\bibfnamefont {Y.}~\bibnamefont {Hai}}, \bibinfo {author} {\bibfnamefont {B.-K.}\ \bibnamefont {Lyu}}, \bibinfo {author} {\bibfnamefont {G.-R.}\ \bibnamefont {Wang}},\ and\ \bibinfo {author} {\bibfnamefont {S.-L.}\ \bibnamefont {Cong}},\ }\bibfield  {title} {\bibinfo {title} {Feshbach resonances of nonzero partial waves at different collision energies},\ }\href {https://doi.org/10.1088/1361-6455/abf8c3} {\bibfield  {journal} {\bibinfo  {journal} {Journal of Physics B: Atomic, Molecular and Optical Physics}\ }\textbf {\bibinfo {volume} {54}},\ \bibinfo {pages} {115201} (\bibinfo {year} {2021})}\BibitemShut {NoStop}%
\bibitem [{\citenamefont {Samuelis}\ \emph {et~al.}(2000)\citenamefont {Samuelis}, \citenamefont {Tiesinga}, \citenamefont {Laue}, \citenamefont {Elbs}, \citenamefont {Kn\"ockel},\ and\ \citenamefont {Tiemann}}]{Samuelis_2000}%
  \BibitemOpen
  \bibfield  {author} {\bibinfo {author} {\bibfnamefont {C.}~\bibnamefont {Samuelis}}, \bibinfo {author} {\bibfnamefont {E.}~\bibnamefont {Tiesinga}}, \bibinfo {author} {\bibfnamefont {T.}~\bibnamefont {Laue}}, \bibinfo {author} {\bibfnamefont {M.}~\bibnamefont {Elbs}}, \bibinfo {author} {\bibfnamefont {H.}~\bibnamefont {Kn\"ockel}},\ and\ \bibinfo {author} {\bibfnamefont {E.}~\bibnamefont {Tiemann}},\ }\bibfield  {title} {\bibinfo {title} {Cold atomic collisions studied by molecular spectroscopy},\ }\href {https://doi.org/10.1103/PhysRevA.63.012710} {\bibfield  {journal} {\bibinfo  {journal} {Phys. Rev. A}\ }\textbf {\bibinfo {volume} {63}},\ \bibinfo {pages} {012710} (\bibinfo {year} {2000})}\BibitemShut {NoStop}%
\bibitem [{\citenamefont {Lysebo}\ and\ \citenamefont {Veseth}(2009)}]{Lysebo_2009}%
  \BibitemOpen
  \bibfield  {author} {\bibinfo {author} {\bibfnamefont {M.}~\bibnamefont {Lysebo}}\ and\ \bibinfo {author} {\bibfnamefont {L.}~\bibnamefont {Veseth}},\ }\bibfield  {title} {\bibinfo {title} {Ab initio calculation of feshbach resonances in cold atomic collisions: $s$- and $p$-wave feshbach resonances in ${^{6}\text{L}\text{i}}_{2}$},\ }\href {https://doi.org/10.1103/PhysRevA.79.062704} {\bibfield  {journal} {\bibinfo  {journal} {Phys. Rev. A}\ }\textbf {\bibinfo {volume} {79}},\ \bibinfo {pages} {062704} (\bibinfo {year} {2009})}\BibitemShut {NoStop}%
\bibitem [{\citenamefont {Weckesser}\ \emph {et~al.}(2021)\citenamefont {Weckesser}, \citenamefont {Thielemann}, \citenamefont {Wiater}, \citenamefont {Wojciechowska}, \citenamefont {Karpa}, \citenamefont {Jachymski}, \citenamefont {Tomza}, \citenamefont {Walker},\ and\ \citenamefont {Schaetz}}]{Weckesser_2021}%
  \BibitemOpen
  \bibfield  {author} {\bibinfo {author} {\bibfnamefont {P.}~\bibnamefont {Weckesser}}, \bibinfo {author} {\bibfnamefont {F.}~\bibnamefont {Thielemann}}, \bibinfo {author} {\bibfnamefont {D.}~\bibnamefont {Wiater}}, \bibinfo {author} {\bibfnamefont {A.}~\bibnamefont {Wojciechowska}}, \bibinfo {author} {\bibfnamefont {L.}~\bibnamefont {Karpa}}, \bibinfo {author} {\bibfnamefont {K.}~\bibnamefont {Jachymski}}, \bibinfo {author} {\bibfnamefont {M.}~\bibnamefont {Tomza}}, \bibinfo {author} {\bibfnamefont {T.}~\bibnamefont {Walker}},\ and\ \bibinfo {author} {\bibfnamefont {T.}~\bibnamefont {Schaetz}},\ }\bibfield  {title} {\bibinfo {title} {Observation of feshbach resonances between a single ion and ultracold atoms},\ }\href {https://doi.org/10.1038/s41586-021-04112-y} {\bibfield  {journal} {\bibinfo  {journal} {Nature}\ }\textbf {\bibinfo {volume} {600}},\ \bibinfo {pages} {429} (\bibinfo {year} {2021})}\BibitemShut {NoStop}%
\bibitem [{\citenamefont {Blech}\ \emph {et~al.}(2020)\citenamefont {Blech}, \citenamefont {Shagam}, \citenamefont {Hoelsch}, \citenamefont {Paliwal}, \citenamefont {Skomorowski}, \citenamefont {Rosenberg}, \citenamefont {Bibelnik}, \citenamefont {Heber}, \citenamefont {Reich}, \citenamefont {Narevicius},\ and\ \citenamefont {Koch}}]{Blech_2020}%
  \BibitemOpen
  \bibfield  {author} {\bibinfo {author} {\bibfnamefont {A.}~\bibnamefont {Blech}}, \bibinfo {author} {\bibfnamefont {Y.}~\bibnamefont {Shagam}}, \bibinfo {author} {\bibfnamefont {N.}~\bibnamefont {Hoelsch}}, \bibinfo {author} {\bibfnamefont {P.}~\bibnamefont {Paliwal}}, \bibinfo {author} {\bibfnamefont {W.}~\bibnamefont {Skomorowski}}, \bibinfo {author} {\bibfnamefont {J.~W.}\ \bibnamefont {Rosenberg}}, \bibinfo {author} {\bibfnamefont {N.}~\bibnamefont {Bibelnik}}, \bibinfo {author} {\bibfnamefont {O.}~\bibnamefont {Heber}}, \bibinfo {author} {\bibfnamefont {D.~M.}\ \bibnamefont {Reich}}, \bibinfo {author} {\bibfnamefont {E.}~\bibnamefont {Narevicius}},\ and\ \bibinfo {author} {\bibfnamefont {C.~P.}\ \bibnamefont {Koch}},\ }\bibfield  {title} {\bibinfo {title} {Phase protection of fano-feshbach resonances},\ }\href {https://doi.org/10.1038/s41467-020-14797-w} {\bibfield  {journal} {\bibinfo  {journal} {Nature Communications}\ }\textbf {\bibinfo {volume} {11}},\ \bibinfo {pages} {999} (\bibinfo {year}
  {2020})}\BibitemShut {NoStop}%
\bibitem [{\citenamefont {Vogels}\ \emph {et~al.}(2018)\citenamefont {Vogels}, \citenamefont {Karman}, \citenamefont {Klos}, \citenamefont {Besemer}, \citenamefont {Onvlee}, \citenamefont {van~der Avoird}, \citenamefont {Groenenboom},\ and\ \citenamefont {van~de Meerakker}}]{Vogels_2018}%
  \BibitemOpen
  \bibfield  {author} {\bibinfo {author} {\bibfnamefont {S.~N.}\ \bibnamefont {Vogels}}, \bibinfo {author} {\bibfnamefont {T.}~\bibnamefont {Karman}}, \bibinfo {author} {\bibfnamefont {J.}~\bibnamefont {Klos}}, \bibinfo {author} {\bibfnamefont {M.}~\bibnamefont {Besemer}}, \bibinfo {author} {\bibfnamefont {J.}~\bibnamefont {Onvlee}}, \bibinfo {author} {\bibfnamefont {A.}~\bibnamefont {van~der Avoird}}, \bibinfo {author} {\bibfnamefont {G.~C.}\ \bibnamefont {Groenenboom}},\ and\ \bibinfo {author} {\bibfnamefont {S.~Y.~T.}\ \bibnamefont {van~de Meerakker}},\ }\bibfield  {title} {\bibinfo {title} {Scattering resonances in bimolecular collisions between no radicals and h2 challenge the theoretical gold standard},\ }\href {https://doi.org/10.1038/s41557-018-0001-3} {\bibfield  {journal} {\bibinfo  {journal} {Nature Chemistry}\ }\textbf {\bibinfo {volume} {10}},\ \bibinfo {pages} {435} (\bibinfo {year} {2018})}\BibitemShut {NoStop}%
\bibitem [{\citenamefont {Naidon}\ and\ \citenamefont {Pricoupenko}(2019)}]{Naidon_2019}%
  \BibitemOpen
  \bibfield  {author} {\bibinfo {author} {\bibfnamefont {P.}~\bibnamefont {Naidon}}\ and\ \bibinfo {author} {\bibfnamefont {L.}~\bibnamefont {Pricoupenko}},\ }\bibfield  {title} {\bibinfo {title} {Width and shift of fano-feshbach resonances for van der waals interactions},\ }\href {https://doi.org/10.1103/PhysRevA.100.042710} {\bibfield  {journal} {\bibinfo  {journal} {Phys. Rev. A}\ }\textbf {\bibinfo {volume} {100}},\ \bibinfo {pages} {042710} (\bibinfo {year} {2019})}\BibitemShut {NoStop}%
\bibitem [{\citenamefont {van Abeelen}\ and\ \citenamefont {Verhaar}(1999)}]{Abeelen_1999}%
  \BibitemOpen
  \bibfield  {author} {\bibinfo {author} {\bibfnamefont {F.~A.}\ \bibnamefont {van Abeelen}}\ and\ \bibinfo {author} {\bibfnamefont {B.~J.}\ \bibnamefont {Verhaar}},\ }\bibfield  {title} {\bibinfo {title} {Time-dependent feshbach resonance scattering and anomalous decay of a na bose-einstein condensate},\ }\href {https://doi.org/10.1103/PhysRevLett.83.1550} {\bibfield  {journal} {\bibinfo  {journal} {Phys. Rev. Lett.}\ }\textbf {\bibinfo {volume} {83}},\ \bibinfo {pages} {1550} (\bibinfo {year} {1999})}\BibitemShut {NoStop}%
\bibitem [{\citenamefont {Paliwal}\ \emph {et~al.}(2021)\citenamefont {Paliwal}, \citenamefont {Blech}, \citenamefont {Koch},\ and\ \citenamefont {Narevicius}}]{Paliwal_2021}%
  \BibitemOpen
  \bibfield  {author} {\bibinfo {author} {\bibfnamefont {P.}~\bibnamefont {Paliwal}}, \bibinfo {author} {\bibfnamefont {A.}~\bibnamefont {Blech}}, \bibinfo {author} {\bibfnamefont {C.~P.}\ \bibnamefont {Koch}},\ and\ \bibinfo {author} {\bibfnamefont {E.}~\bibnamefont {Narevicius}},\ }\bibfield  {title} {\bibinfo {title} {Fano interference in quantum resonances from angle-resolved elastic scattering},\ }\href {https://doi.org/10.1038/s41467-021-27556-2} {\bibfield  {journal} {\bibinfo  {journal} {Nature Communications}\ }\textbf {\bibinfo {volume} {12}},\ \bibinfo {pages} {7249} (\bibinfo {year} {2021})}\BibitemShut {NoStop}%
\bibitem [{\citenamefont {de~Jongh}\ \emph {et~al.}(2020)\citenamefont {de~Jongh}, \citenamefont {Besemer}, \citenamefont {Shuai}, \citenamefont {Karman}, \citenamefont {van~der Avoird}, \citenamefont {Groenenboom},\ and\ \citenamefont {van~de Meerakker}}]{Jongh_2020}%
  \BibitemOpen
  \bibfield  {author} {\bibinfo {author} {\bibfnamefont {T.}~\bibnamefont {de~Jongh}}, \bibinfo {author} {\bibfnamefont {M.}~\bibnamefont {Besemer}}, \bibinfo {author} {\bibfnamefont {Q.}~\bibnamefont {Shuai}}, \bibinfo {author} {\bibfnamefont {T.}~\bibnamefont {Karman}}, \bibinfo {author} {\bibfnamefont {A.}~\bibnamefont {van~der Avoird}}, \bibinfo {author} {\bibfnamefont {G.~C.}\ \bibnamefont {Groenenboom}},\ and\ \bibinfo {author} {\bibfnamefont {S.~Y.~T.}\ \bibnamefont {van~de Meerakker}},\ }\bibfield  {title} {\bibinfo {title} {Imaging the onset of the resonance regime in low-energy no-he collisions},\ }\href {https://doi.org/10.1126/science.aba3990} {\bibfield  {journal} {\bibinfo  {journal} {Science}\ }\textbf {\bibinfo {volume} {368}},\ \bibinfo {pages} {626} (\bibinfo {year} {2020})}\BibitemShut {NoStop}%
\bibitem [{\citenamefont {Blatt}\ and\ \citenamefont {Biedenharn}(1952)}]{Blatt_1952}%
  \BibitemOpen
  \bibfield  {author} {\bibinfo {author} {\bibfnamefont {J.~M.}\ \bibnamefont {Blatt}}\ and\ \bibinfo {author} {\bibfnamefont {L.~C.}\ \bibnamefont {Biedenharn}},\ }\bibfield  {title} {\bibinfo {title} {The angular distribution of scattering and reaction cross sections},\ }\href {https://doi.org/10.1103/RevModPhys.24.258} {\bibfield  {journal} {\bibinfo  {journal} {Rev. Mod. Phys.}\ }\textbf {\bibinfo {volume} {24}},\ \bibinfo {pages} {258} (\bibinfo {year} {1952})}\BibitemShut {NoStop}%
\bibitem [{\citenamefont {Rose}(1957)}]{Rose_1957}%
  \BibitemOpen
  \bibfield  {author} {\bibinfo {author} {\bibfnamefont {M.~E.}\ \bibnamefont {Rose}},\ }\href@noop {} {\emph {\bibinfo {title} {Elementary Theory of Angular Momentum}}}\ (\bibinfo  {publisher} {New York: Wiley},\ \bibinfo {year} {1957})\BibitemShut {NoStop}%
\bibitem [{\citenamefont {Palov}\ and\ \citenamefont {Balint-Kurti}(2021)}]{PALOV2021107895}%
  \BibitemOpen
  \bibfield  {author} {\bibinfo {author} {\bibfnamefont {A.}~\bibnamefont {Palov}}\ and\ \bibinfo {author} {\bibfnamefont {G.}~\bibnamefont {Balint-Kurti}},\ }\bibfield  {title} {\bibinfo {title} {Vpa: Computer program for the computation of the phase shift in atom–atom potential scattering using the variable phase approach},\ }\href {https://doi.org/https://doi.org/10.1016/j.cpc.2021.107895} {\bibfield  {journal} {\bibinfo  {journal} {Computer Physics Communications}\ }\textbf {\bibinfo {volume} {263}},\ \bibinfo {pages} {107895} (\bibinfo {year} {2021})}\BibitemShut {NoStop}%
\bibitem [{\citenamefont {{Srinivasa Rao}}\ and\ \citenamefont {Venkatesh}(1978)}]{Rao1978}%
  \BibitemOpen
  \bibfield  {author} {\bibinfo {author} {\bibfnamefont {K.}~\bibnamefont {{Srinivasa Rao}}}\ and\ \bibinfo {author} {\bibfnamefont {K.}~\bibnamefont {Venkatesh}},\ }\bibfield  {title} {\bibinfo {title} {New fortran programs for angular momentum coefficients},\ }\href@noop {} {\bibfield  {journal} {\bibinfo  {journal} {Computer Physics Communications}\ }\textbf {\bibinfo {volume} {15}},\ \bibinfo {pages} {227} (\bibinfo {year} {1978})}\BibitemShut {NoStop}%
\end{thebibliography}

\providecommand{\noopsort}[1]{}\providecommand{\singleletter}[1]{#1}%

\end{document}